\documentclass[lettersize,journal]{IEEEtran}
\IEEEoverridecommandlockouts

\usepackage{generic}
\usepackage{cite}
\usepackage{amssymb,amsfonts}
\usepackage{pifont}

\usepackage[pdftex]{graphicx}
\usepackage{algorithmic}
\usepackage{dblfloatfix}
\usepackage{textcomp}
\usepackage{caption}
\usepackage{subscript}
\usepackage{lipsum}
\usepackage[fleqn]{amsmath}
\usepackage{microtype}
\usepackage{nccmath, mathtools}

\usepackage{subcaption}
\usepackage{siunitx}
\DeclareSIUnit\million{mil.}
\usepackage{numprint}
\npdecimalsign{.}
\usepackage{booktabs}

\usepackage{makecell}

\usepackage{nicefrac}

\usepackage{graphicx}
\usepackage{tikz}
\usetikzlibrary{calc,fadings}
\usetikzlibrary{shapes.geometric}
\usepackage{circuitikz}

\usepackage{pgfplots}
\pgfplotsset{compat=1.18}
\usepgfplotslibrary{groupplots,colormaps}

\usepackage{pgfplotstable}

\usepackage{svg}

\usepackage{xparse}
\ExplSyntaxOn
\DeclareExpandableDocumentCommand{\convertlen}{ O{cm} m }
{
	\dim_to_decimal_in_unit:nn { #2 } { 1 #1 } cm
}
\ExplSyntaxOff

\usepackage{xspace}
\xspaceaddexceptions{)}
\usepackage{glossaries}

\newglossaryentry{vdd}{
	name={\ensuremath{\text{{VDD}}}},
	description={supply voltage},
	symbol=Vdd,
	sort=vdd}
\newglossaryentry{vss}{
	name={\ensuremath{\text{{VSS}}}},
	description={supply voltage},
	symbol=Vss,
	sort=vss}
\newglossaryentry{vth}{
	name={\ensuremath{V_{\text{TH}}}},
	description={threshold voltage},
	symbol=Vth,
	sort=vth}
\newglossaryentry{gnd}{
	name={GND},
	description={El. ground},
	symbol=GND}

\newacronym{nvm}{NVM}{non-volatile memory}

\newacronym{dnn}{DNN}{deep neural network}
\newacronym{ai}{AI}{artificial intelligence}
\newacronym{iot}{IoT}{Internet-of-Things}

\newacronym{sram}{SRAM}{Static random-access memory}
\newacronym{nvsram}{nvSRAM}{non-volatile SRAM}
\newacronym{ml}{ML}{matchline}
\newacronym{sl}{SL}{signal line}
\newacronym{wl}{WL}{wordline}
\newacronym{bl}{BL}{bitline}
\newacronym{blb}{BLB}{bitline bar}
\newacronym{slb}{SLB}{signal line bar}

\newacronym{snm}{SNM}{Static Noise Margin}
\newacronym{hsnm}{HSNM}{hold SNM}
\newacronym{rsnm}{RSNM}{read SNM}
\newacronym{wsnm}{WSNM}{write SNM}

\newacronym{tcad}{TCAD}{technology CAD}

\newacronym{ss}{SS}{sub-threshold slope}

\newacronym{fdsoi}{FDSOI}{Fully-Depleted Silicon on Insulator}
\newacronym{fefet}{FeFET}{Ferroelectric FET}
\newacronym{fe}{FE}{ferroelectric}
\newacronym{mlc}{MLC}{multi-level cell}

\newacronym{ds}{DS}{Domain Stochasticity}

\newacronym{dtd}{DTD}{Device-to-Device}
\newacronym{ctc}{CTC}{Cycle-to-Cycle}
\newacronym{mw}{MW}{memory window}

\newacronym{stt}{STT}{Spin Transfer Torque}
\newacronym{rram}{ReRAM}{Resistive RAM}
\newacronym{mtj}{MTJ}{magnetic tunnel junctions}
\newacronym{fecap}{FeCAP}{ferroelectric capacitor}

\newacronym{pvt}{PVT}{process, voltage, and temperatur}

\newacronym{dut}{DUT}{device under test}
\newacronym{smu}{SMU}{source measurement unit}

\glsunset{slb}
\glsunset{blb}

\newcommand{\nvm}{\gls{nvm}\xspace}

\newcommand{\sram}{\gls{sram}\xspace}
\newcommand{\nvsram}{\gls{nvsram}\xspace}
\newcommand{\nvsrams}{\glspl{nvsram}\xspace}

\newcommand{\bitline}{\gls{bl}\xspace}
\newcommand{\blb}{\gls{blb}\xspace}
\newcommand{\wordline}{\gls{wl}\xspace}
\newcommand{\signalline}{\gls{sl}\xspace}

\newcommand{\stateq}{\ensuremath{\text{Q}}\xspace}
\newcommand{\stateqb}{\ensuremath{\text{QB}}\xspace}

\newcommand{\snm}{\gls{snm}\xspace}
\newcommand{\hsnm}{\gls{hsnm}\xspace}
\newcommand{\rsnm}{\gls{rsnm}\xspace}
\newcommand{\wsnm}{\gls{wsnm}\xspace}

\newcommand{\fdsoi}{\gls{fdsoi}\xspace}
\newcommand{\fefet}{\gls{fefet}\xspace}
\newcommand{\fefets}{\glspl{fefet}\xspace}
\newcommand{\nfefet}{n-\gls{fefet}\xspace}
\newcommand{\nfefets}{n-\glspl{fefet}\xspace}
\newcommand{\pfefet}{p-\gls{fefet}\xspace}
\newcommand{\pfefets}{p-\glspl{fefet}\xspace}
\newcommand{\fe}{\gls{fe}\xspace}

\newcommand{\memwindow}{\gls{mw}\xspace}

\newcommand{\dut}{\gls{dut}\xspace}

\newcommand{\smus}{\glspl{smu}\xspace}

\newcommand{\vdd}{\gls{vdd}\xspace}

\newcommand{\vth}{\gls{vth}\xspace}
\newcommand{\vds}{\ensuremath{V_{\text{DS}}}\xspace}
\newcommand{\vd}{\ensuremath{V_{\text{D}}}\xspace}
\newcommand{\vgs}{\ensuremath{V_{\text{G}}}\xspace}
\newcommand{\vg}{\ensuremath{V_{\text{G}}}\xspace}
\newcommand{\vs}{\ensuremath{V_{\text{S}}}\xspace}

\newcommand{\lvt}{LVT\xspace}
\newcommand{\hvt}{HVT\xspace}

\newcommand{\ion}{\ensuremath{I_{\text{ON}}}\xspace}
\newcommand{\ioff}{\ensuremath{I_{\text{OFF}}}\xspace}
\newcommand{\ids}{\ensuremath{I_{\text{D}}}\xspace}
\newcommand{\idvg}{\ensuremath{\ids\text{-}\vgs}\xspace}

\newcommand{\code}[1]{\texttt{#1}}
\newcommand{\chdim}[2]{\ensuremath{#1 \times #2}}
\newcommand{\chdimsqr}[1]{\chdim{#1}{#1}}

\newcommand{\logicone}{\code{1}\xspace}
\newcommand{\logiczero}{\code{0}\xspace}

\newcommand{\hfo}{\ensuremath{\text{HfO}_{2}}\xspace}

\newcommand{\mosfet}{MOSFET\xspace}
\newcommand{\spice}{SPICE\xspace}
\newcommand{\cmos}{CMOS\xspace}

\newcommand{\pmos}{PMOS\xspace}
\newcommand{\nmos}{NMOS\xspace}
\usepackage{csquotes}
\usepackage[hidelinks]{hyperref}
\usepackage{orcidlink}

\newcommand{\cmark}{\ding{51}}%
\newcommand{\xmark}{\ding{55}}%

\definecolor{lvtcolor}{rgb}{1,0,0}
\definecolor{hvtcolor}{rgb}{0,0,1}

\colorlet{mycyan}{teal}
\definecolor{mygreen}{RGB}{58, 90, 64}

\definecolor{fefetcolor}{RGB}{255, 183, 3}

\def\nobreakbefore{%
  \relax\ifvmode\else
    \ifhmode
      \ifdim\lastskip > 0pt\relax
        \unskip\nobreakspace
      \fi
    \fi
  \fi
}
\let\oldcite\cite
\renewcommand\cite{\nobreakbefore\oldcite}

\usepackage{cleveref}

\crefname{enumi}{Step}{Steps}
\crefname{section}{Sec.}{Sec.}
\crefname{subsection}{Sec.}{Sec.}
\crefname{figure}{Fig.}{Fig.}
\crefname{algocf}{Algo.}{Algo.}
\crefname{algorithm}{Algo.}{Algo.}
\crefname{algocf}{Algo.}{Algo.}
\crefname{equation}{Eq.}{Eq.}
\crefname{eqnarray}{Eq.}{Eq.}
\crefname{appendix}{Sec.}{Sec.}

\crefname{table}{Table}{Tables}

\begin{document}

\title{First Demonstration of 28 nm Fabricated FeFET-Based Nonvolatile 6T SRAM}

\author{Albi Mema$^{\orcidlink{0000-0001-7841-1975}}$,~\IEEEmembership{Member,~IEEE},
        Simon Thomann$^{\orcidlink{0000-0002-7902-9353}}$,~\IEEEmembership{Member,~IEEE},
        Narendra Singh Dhakad$^{\orcidlink{0000-0003-2848-1785}}$,\\
        and Hussam~Amrouch$^{\orcidlink{0000-0002-5649-3102}}$,~\IEEEmembership{Member,~IEEE}
    \thanks{This work was supported by the German Research Foundation (DFG) under grant AM 534/8-1 (DeCOPE, project: 547377347). \emph{(Corresponding author: Hussam Amrouch)}}
    
	\thanks{%
		Albi Mema, Simon Thomann, Narendra Singh Dhakad, and Hussam Amrouch are with the Technical University of Munich; TUM School of Computation, Information and Technology; Chair of AI Processor Design; Munich Institute of Robotics and Machine Intelligence (MIRMI), Munich, Germany. The work of Narendra S. Dhakad was solely done during his employment (2024-2025) at the TUM Chair of AI Processor Design. E-mail: \{a.mema, s.thomann, amrouch\}@tum.de
	}%
}

\maketitle

\begin{abstract}
With the staggering increase of edge compute applications like Internet-of-Things (IoT) and artificial intelligence (AI), the demand for fast, energy-efficient on-chip memory is growing.
While the fast and mature static random-access memory (SRAM) technology is the standard choice, its volatility requires a constant supply voltage to operate and store data.
Especially in edge AI and IoT devices that often idle, the leakage power consumes a significant portion of the constrained power budget.
For this, emerging non-volatile memory (NVM) technologies such as Resistive RAM and ferroelectric FET (FeFET) offer zero-standby power consumption but suffer from integration and performance tradeoffs.
To harness the benefits of the different technologies, hybrid architectures have been proposed, combining SRAM with NVM devices.
This work proposes a hybrid non-volatile SRAM (nvSRAM) architecture based on recently demonstrated PMOS FeFETs (p-FeFETs).
By replacing the two \pmos pull-up transistors with p-FeFETs, we achieve non-volatility without additional transistors.
The design supports seamless power-down and restore operation, thus eliminating standby leakage.
SPICE simulations in a commercial 28 nm technology show read latency comparable to conventional SRAM, and on-silicon measurements show robust restore behavior.
With this, we are the first to demonstrate a fabricated 6T nvSRAM cell design.
The resulting cell achieves an area footprint of \qty{0.99}{\micro\meter\squared}.
The read path remains identical to baseline SRAM, enabling high-speed operation while being non-volatile, making it ideal for IoT and edge systems.

\end{abstract}

\begin{IEEEkeywords}
nvSRAM, Ferroelectric, FeFET, Non-volatile, Ultra-low-power
\end{IEEEkeywords}
\vspace{-0.4cm}

\section{Introduction}
\label{sec:introduction}

The proliferation of edge computing, \gls{iot}, and \gls{ai} has intensified the demand for fast, energy-efficient on-chip memory.
\Gls{sram} is the standard choice for these applications due to its high speed, reliability, and technological maturity.
In modern \gls{ai} accelerators \sram area constitutes up to \qty{80}{\percent} of the chip area \cite{IMC_AP, Nature2025, CONV_SRAM, challanges}.
However, \sram is volatile by design and requires a continuous power supply to retain data.
In edge \gls{ai} and \gls{iot} devices that remain idle for a significant period and operate in a power-constrained environment, this results in substantial leakage power consumption \cite{Ref_Power1, Ref_Power2}. 

To mitigate these drawbacks, emerging \nvm technologies such as \gls{rram} \cite{7T1R, 7T2M, 4nvSRAM, 6T2R2S, Sharma_RRAM} and \fefets \cite{FeFET1, FeFET2, FeFET3, FeFET4, FeFET5, FeFET6} have been explored.
While these technologies offer zero power consumption to retain data, they often suffer from reliability concerns, high error rates, relatively long write and read latencies, and, in the case of \gls{rram}, considerably higher write power consumption \cite{8T2F}.
Thus, this limits their adoption into \gls{ai}-centric applications. 

Consequently, hybrid memory architectures have recently been gaining traction.
These architectures leverage both the strengths of \sram and non-volatile memory approaches.
\gls{rram}~\cite{7T1R, 7T2M, 4nvSRAM, 6T2R2S}, \fefets \cite{8T2F, 7T1F, UCB}, \glspl{fecap} \cite{6T2C, 6T4C} and \gls{mtj} \cite{MTJ, MTJ8T, MEFET} are some of the non-volatile devices integrated within the \sram cell.
Among these, \fefets and \glspl{fecap} are particularly promising due to their \cmos compatibility and scalability \cite{trentzsch2016hkmg28nm_fefet}.
However, existing \fefet-based designs face significant trade-offs.
For instance, 6T2C architectures~\cite{6T2C} are highly sensitive to transistor mismatch, while 6T4C designs~\cite{6T4C} incur a \qty{46}{\percent} area overhead.
Other recent \nvsrams utilize \nmos \fefets (\nfefets) as auxiliary transistors~\cite{8T2F, 7T1F, UCB}, which increases the silicon footprint and complexity of the memory array.

\begin{figure*}[t]
    \vspace{-0.3cm}
    \centering

    \begin{subfigure}[c]{\columnwidth}
        \centering
        \includegraphics{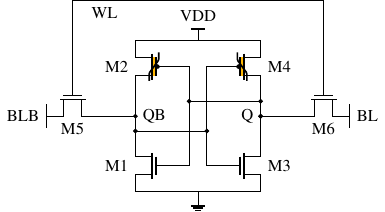}
        \caption{Proposed \nvsram}
        \label{fig:nvsram_circuit}
    \end{subfigure}%
    \begin{subfigure}[c]{\columnwidth}
        \centering
        \includegraphics{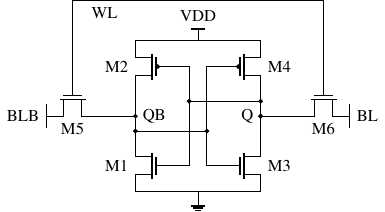}
        \caption{Baseline 6T \sram}
        \label{fig:baseline_sram_circuit}
    \end{subfigure}%
    
    \vspace{0.25cm}
    
    \begin{subfigure}[c]{\columnwidth}
        \centering
        \includegraphics{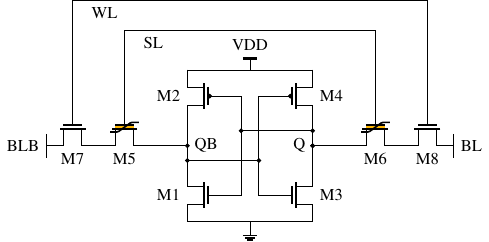}
        \caption{8T \nvsram \cite{UCB}}
        \label{fig:8T_nvsram_circuit}
    \end{subfigure}%
    \begin{subfigure}[c]{\columnwidth}
        \centering
        \includegraphics{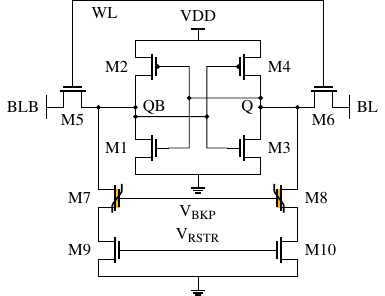}
        \caption{10T \nvsram \cite{8T2F}}
        \label{fig:10T_nvsram_circuit}
    \end{subfigure}%
    
    \caption{Circuit diagram of (a) our proposed 6T \nvsram, (b) baseline 6T \sram, (c) 8T \nvsram \cite{UCB}, and (d) 10T \nvsram \cite{8T2F}.
    Here, the transistors highlighted in orange are \fefets.}
    \label{fig:Intro_Compare}

\end{figure*}

\begin{table*}[t]
    \centering
    \caption{Performance Matrix Comparison of \sram, Non-Volatile Memories and Our Proposed Hybrid Memory}
    \renewcommand{\arraystretch}{1.2}
    \begin{tabular}{@{}l l l l@{}}
    \toprule
    \textbf{Metric}             & \textbf{Baseline}                                         & \textbf{Non-Volatile Memories}    & \textbf{Proposed Hybrid Memory}      \\
    \midrule
    \textbf{Non-Volatility}     & \xmark                   & \cmark                               & \cmark     \\ 
    \textbf{Area}               & \qtyrange[range-units=single, range-phrase=--]{0.08}{0.2}{\micro\meter\squared} \cite{SRAM_survey, sram_data_1}   & \qtyrange[range-units=single, range-phrase=--]{0.076}{1.13}{\micro\meter\squared} \cite{Memories_matrix} & \qty{0.992}{um} \Cref{fig:nvsram_layout}   \\ 
    \textbf{Idle Power}         & \qtyrange[range-units=single, range-phrase=--]{1}{50}{\nano\watt\per\text{cell}} \cite{SRAM_survey, sram_data_1}  & No power (OFF) \cite{nvm_review} & No power (OFF) \\ 
    \textbf{Read Latency}       & \qty{1}{ps} -- \qty{10}{ns} \cite{SRAM_survey, sram_data_1}                                                       & \qtyrange[range-units=single, range-phrase=--]{2}{40}{ns} \cite{Memories_matrix} & \qty{72}{ps}  \Cref{tab:nvsram_to_baseline_comp}    \\ 
    \textbf{Write Latency}      & \qty{10}{ps} -- \qty{20}{ns} \cite{SRAM_survey, sram_data_1}                                                      & \qty{1}{ns} -- \qty{12}{us}  \cite{Memories_matrix} & \qty{10}{ns} \cite{dunkel2017fefet_22nm, mulasmanovic2019gf_fefet_endurance} \\ 
    \textbf{Read energy}        & \qty{1}{fJ} -- \qty{1}{pJ} \cite{SRAM_survey, sram_data_2}                                                        & \qtyrange[range-units=single, range-phrase=--]{1}{1.2}{pJ} \cite{Memories_matrix} & \qty{1}{fJ} -- \qty{1}{pJ} \cite{SRAM_survey, sram_data_2} \\ 
    \textbf{Write energy}       & \qty{5}{fJ} -- \qty{5}{pJ} \cite{SRAM_survey, sram_data_2}                                                        & \qty{5}{fJ} -- \qty{12}{pJ}  \cite{Memories_matrix} & \qtyrange[range-units=single, range-phrase=--]{0.1}{2}{pJ} \cite{fefet_comparative_analysis} \\ 
    \textbf{Data Retention}     & None  &  \qtyrange[range-units=single, range-phrase=--]{10}{121}{\text{years}} \cite{Memories_matrix}             & \num{10} years  \cite{trentzsch2016hkmg28nm_fefet}   \\ 
    \bottomrule
    \end{tabular}
    \label{tab:performance_comparison}
\end{table*}

To address these limitations, we propose a novel hybrid \nvsram architecture, utilizing \pmos \fefets (\pfefets) into a 6T \sram architecture shown in \cref{fig:nvsram_circuit}.
By replacing the standard \pmos pull-up transistors with \pfefets, we achieve non-volatility without adding extra transistors or capacitors.
This design enables high-speed operations and seamless power-down states with minimal area overhead, making it ideal for energy-constrained \gls{iot} and edge systems.
This approach is only possible with the recent fabrication of \pfefets.
During read operations, the bitlines are precharged to \vdd, and the discharge path is dominated by the pull-down \nmos transistor to ground.
As a result, the load \pfefets do not participate in the read current path, preserving the read performance of the \sram.
Due to its non-volatile nature, the proposed design can be effectively powered down during long idle states without losing the stored information in the bitcells, thereby reducing power consumption from the memory array to zero. 

Alongside our proposed design, \Cref{fig:Intro_Compare} shows the schematic of the conventional 6T \sram and state-of-the-art 8T and 10T \nvsram architectures \cite{UCB, 8T2F}.
Further, \cref{tab:performance_comparison} compares the performance metrics for the conventional \sram, non-volatile memories, and proposed hybrid \nvsram architecture.

\newpage
\noindent
\textbf{Our novel contributions within this paper are as follows:}\\
(1) Novel Topology: We introduce the first \nvsram design that utilizes \pfefets within the 6T \sram structure to replace standard \pmos transistors.\\
\noindent
(2) Area Efficiency: The proposed architecture achieves non-volatility without increasing the device count.\\
\noindent
(3) Experimental Validation: We validate the design's robustness against variability using extensive SPICE simulations based on a commercial \qty{28}{nm} PDK.
We are the \emph{first} to validate the proposed design of a \fefet-based 6T \nvsram cell in silicon.
We demonstrate this by performing measurements on a taped-out chip in the \qty{28}{nm} technology, which supports \fefet transistors.

\section{Background}
\label{sec fdsoi basics}

\begin{table}[t]
    \centering
    \caption{Literature survey of \nvsram designs.}
    \label{tab:nvSRAM_summary}
    \renewcommand{\arraystretch}{1.2}
    \resizebox{\columnwidth}{!}{
    \begin{tabular}{@{}ccccc@{}}
        \toprule
        Reference & Simulated & Fabricated & Technology & Cell Configuration \\
        \midrule
        Our & \cmark & \cmark & \fefet & 6T \\
        \midrule
        \cite{ftj_fabricated_nv} & \cmark & \cmark & FTJ & 6T2C \\
        \cite{liu2025_fecap_beol_3d_nvsram_cell} & \cmark & \cmark & FeCAP + TFT & 4T2C \\
        \cite{kobayashi_fecap_nvsram_demo} & \cmark & \cmark & FeCAP & Not disclosed \\
        \cite{rram_nvsram_fabricated} & \cmark & \cmark & RRAM & 12T2R \\
        \cite{UCB} & \cmark & \xmark & \fefet & 8T \\
        \cite{8T2F} & \cmark & \xmark & \fefet & 10T \\
        \cite{8t_fefet_simulated} & \cmark & \xmark & \fefet & 8T \\
        \cite{fefet_simulated_2} & \cmark & \xmark & R\fefet & 3T-R \\
        \cite{rram_nvsram_sim1} & \cmark & \xmark & RRAM & 8T2R \\
        \cite{rram_nvsram_sim2} & \cmark & \xmark & RRAM & 7T1R \\
        \cite{rram_nvsram_sim3} & \cmark & \xmark & RRAM & 7T1R \\
        \cite{rram_nvsram_sim4} & \cmark & \xmark & RRAM & 8T1R \\
        \cite{rram_nvsram_sim5} & \cmark & \xmark & RRAM & 9T1R \\
        \cite{mtj_nvsram_sim} & \cmark & \xmark & MTJ & 8T2R \\
        \cite{memristor_nvsram_sim1} & \cmark & \xmark & Memristor & 8T2R \\
        \cite{memristor_nvsram_sim2} & \cmark & \xmark & Memristor & 8T1R \\
        \bottomrule
    \end{tabular}
    }
\end{table}

\begin{figure}
    \centering
    \footnotesize{}
    \begin{subfigure}[c]{0.5\columnwidth}
        \centering
        \includegraphics{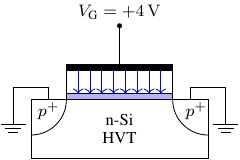}
        \caption{Positive voltage gate write.}
        \label{fig:fefet_pos_gate_write}
    \end{subfigure}%
    \begin{subfigure}[c]{0.5\columnwidth}
        \centering
        \includegraphics{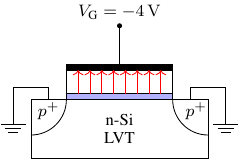}
        \caption{Negative voltage gate write.}
        \label{fig:fefet_neg_gate_write}
    \end{subfigure}%
    \vspace{0.2cm}

    \begin{subfigure}[c]{0.5\columnwidth}
        \centering
        \includegraphics{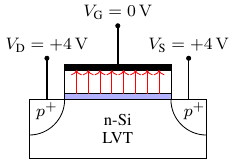}
        \caption{Source and drain write.}
        \label{fig:fefet_source_drain_write}
    \end{subfigure}%
    \begin{subfigure}[c]{0.5\columnwidth}
        \centering
        \includegraphics{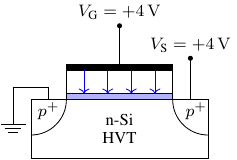}
        \caption{Gate-drain write.}
        \label{fig:fefet_drain_write}
    \end{subfigure}%
    \caption{(a) Conventional high positive gate voltage write.
    The high potential difference from the gate to the grounded source and drain polarizes the ferroelectric layer to \hvt.
    (b) Conventional high negative gate voltage write. This voltage polarizes the ferroelectric to \lvt.
    (c) Source-drain write is employed by the proposed design.
    This scheme omits the need for negative voltages.
    (d) Gate-drain write employed by the proposed design.
    The polarization mainly occurs on the gate-to-drain side of the transistor.
    On average, the ferroelectric layer polarizes to \hvt.}
    \label{fig:fefet_write_schemes}
    \vspace{-0.2cm}
\end{figure}

\subsection{Baseline 6T SRAM}
\label{sec:baseline_6t}
The 6T \sram cell, shown in \Cref{fig:baseline_sram_circuit}, is composed of two cross-coupled \cmos inverters (M1-M4) that form a bistable latch for data storage, and two \nmos access transistors (M5, M6) that connect the internal nodes (\stateq, \stateqb) with the \acrlongpl{bl} (\glsunset{bl}\acrshort{bl}, \acrshort{blb}) when the \wordline is high.

The stored logic states are:
\begin{itemize}
    \item \textbf{Logic \logicone:} $\stateq = \vdd$ \& $\stateqb = \qty{0}{V}$
    \item \textbf{Logic \logiczero:} $\stateq = \qty{0}{V}$ \& $\stateqb = \vdd$
\end{itemize}

\textbf{Write operation:} During the write operations, the bitlines are driven to complementary voltage levels based on the input data to be written.
When \wordline is asserted, the access transistors allow the potential in the \bitline/\blb to be transferred to the \stateq/\stateqb internal nodes, thus overpowering the positive feedback loop of the inverters.
This forces the new value to be stored and preserved even after the \wordline is deasserted.

\textbf{Read operation:} Both bitlines are precharged to \vdd.
When the \wordline is asserted, the internal node \stateq/\stateqb holding \qty{0}{V} discharges the respective bitline, creating a voltage differential on the bitlines.
A sense amplifier detects this differential to determine the stored bit, then the \wordline is deasserted.

\textbf{Hold operation:} With the \wordline deasserted and \vdd enabled, the cell retains its data through the positive feedback.
In this state, the power consumption is dominated by leakage, while the bitlines remain precharged to \vdd.

The baseline 6T \sram provides high speed and reliability, but suffers from 1. \textbf{volatility}, as data is lost when power is removed, and 2. \textbf{idle power consumption}, due to leakage under continuous power supply.

\subsection{State-of-the-art nvSRAM Design}
\label{sec:sota}
To overcome the challenge of 6T \sram described above, integrating non-volatile memory elements (\gls{rram}, \gls{mtj}, \gls{fecap}, and \fefet) into \sram is a common approach to enable non-volatility while preserving \sram-level performance.
This work focuses on \fefet-based \nvsram due to their \cmos compatibility, faster switching speeds, lower write energy~\cite{UCB}, and higher \ion/\ioff ratio.
Additionally, \fefets offer better energy efficiency compared to \gls{mtj}- and \gls{rram}-based designs~\cite{FeFET_EnergyEffecient}.

\begin{figure}[t]
    \centering
    \includegraphics{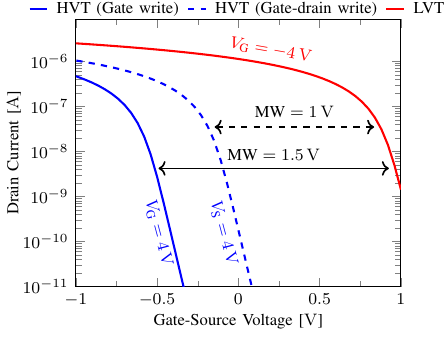}
    \caption{Comparison of \idvg characteristics between the conventional gate programming scheme and drain programming scheme.
    The \acrfull{mw} measures the difference between the \lvt and \hvt states.}
    \label{fig:fefet_write_iv_comp}
    \vspace{-0.2cm}
\end{figure}

State-of-the-art \fefet-based \nvsrams \cite{UCB, 8T2F} shown in \Cref{fig:8T_nvsram_circuit,fig:10T_nvsram_circuit}, achieve non-volatility by adding \fefet-based backup and restore circuitry at the storage nodes \stateq/\stateqb.
Li et al. \cite{8T2F} proposed an \nvsram as shown in \Cref{fig:10T_nvsram_circuit}.
While effective at achieving non-volatility, their proposed design requires four additional transistors.
This, in turn, introduces higher parasitic capacitances on the internal storage nodes, which degrade both read and write access times.
The alternative design from Wang et al. \cite{7T1F} reduces the additional transistor overhead to 2, but it requires independent power-supply control for the cross-coupled inverters and an extra control signal, increasing peripheral complexity.
You et al. in \cite{UCB}, shown in \Cref{fig:8T_nvsram_circuit}, proposed a simplified architecture by inserting two \fefets between \stateq/\stateqb and the access transistors.
Although this approach also incurs minimal transistor-count overhead, it relies on the additional \signalline and gate-write scheme, which adds to the complexity and area overhead.
In \cref{tab:nvSRAM_summary} we list proposed \nvsram cell designs from the literature and detail the used technology, cell configuration, and whether they were validated through fabrication or only via simulation.

\subsection{FeFET as an Emerging Non-Volatile Memory}
\label{sec:fefet_background}

Structurally, the concept of a \fefet is based on a regular \mosfet.
By adding a layer of \fe material to the gate stack, the transistor characteristic can be shifted through the polarization state of the \fe layer.
The polarization of the \fe layer in a \fefet is changed by applying a strong voltage bias, e.g., \qty{\pm 4}{V}, to its gate (see \cref{fig:fefet_pos_gate_write,fig:fefet_neg_gate_write}).
This flips the domains inside the \fe layer to align with the electric field established by the gate potential.
The polarization interacts with the perceived gate voltage in the channel, amplifying or dampening it depending on its direction.
This shifts the transfer characteristic of the \fefet, which is usually described as a shift in threshold voltage (\vth) and ultimately changes the resistance at a given gate voltage.
Using the \emph{two} distinct polarization states allows for storing information in the \fe layer, labeled \lvt and \hvt for the saturated positive/negative directions, respectively.
The opening between the \lvt and \hvt states in the \idvg characteristic is called \emph{\memwindow} and used as a metric for the separation of the two states.
Additionally, \fefet provides non-volatile information storage, as the polarization state persists even when power is cut, making \fefet an \nvm.
With the discovery of doped \hfo-based \fe materials, a thin layer (e.g., \qty{10}{nm}) is sufficient, enabling further scalability of \fefet \cite{firstFefet2011boeschke}.
Following the introduction of doped \hfo-based \fefets, GlobalFoundries demonstrated the fabrication of \qty{22}{nm} \fdsoi \fefets, emphasizing the scalability \cite{dunkel2017fefet_22nm}.
The recent demonstration of \pmos-based \fefets enables new design opportunities for circuits using \fefet technology \cite{kleimaier2021p_fefet}.

The multi-domain structure of the \fe layer allows for a partial polarization of the \fefet.
In addition to the classical gate-driven polarization scheme as shown in \cref{fig:fefet_pos_gate_write,fig:fefet_neg_gate_write}, an electric field can also be established by biasing the source and/or drain terminal \cite{meihar2024mirror_bit_fefet_cell,wang2020fefet_drain_erase}.
Essential for the polarization process is a strong enough potential drop \emph{across} the \fe layer that forces the domains to align their polarization direction.
How this potential difference is established is up to the designer and the circuit's constraints.
\cref{fig:fefet_source_drain_write} shows the biasing using the source \emph{and} drain terminal with \qty[retain-explicit-plus]{+4}{V} and a grounded gate.
This achieves the same direction and electric field strength as the gate biasing with \qty{-4}{V} (see \cref{fig:fefet_neg_gate_write}), thus polarizing the transistor to \lvt, and can be considered equivalent.
Using only the drain (or the source) terminal to establish the electric field, as shown in \cref{fig:fefet_drain_write}, will polarize the \fe portion close to the drain terminal the strongest and weaken towards the source terminal.
This will result in an overall weaker \hvt polarization state as shown in \cref{fig:fefet_write_iv_comp} \cite{meihar2024mirror_bit_fefet_cell}.

\section{Design of Our Proposed nvSRAM Cell}
\label{sec:design}

To overcome the complexity and performance limitations of existing non-volatile memory architectures, we propose a hybrid \nvsram architecture that preserves the baseline 6T SRAM topology.
Our design, as shown in \cref{fig:nvsram_circuit}, integrates \pfefets as the load transistors (M2 and M4).
The proposed design preserves the peripheral architecture of the baseline 6T SRAM while enabling non-volatile operation.
Read and hold modes are identical to the baseline design, ensuring that the critical read path, sense circuitry, and timing remain unchanged.
Non-volatility is achieved by programming the load \pfefets, which require elevated supply voltage.
This elevated supply voltage can be implemented using standard multiplexing without the need for negative voltages and non-standard drivers.
Because the read current path is unchanged and the load devices do not directly participate in the bitline's discharge during read, the proposed design maintains conventional read performance while avoiding complex peripheral circuitry.
This compatibility allows the proposed \nvsram to integrate seamlessly into existing SRAM macro design flows.

\subsection{Write Operation}
The simplicity of the peripheral interface arises from the fact that the logical write sequence is identical to the baseline 6T SRAM, with only the amplitude temporarily increased.
The bitlines (\bitline and \blb) are driven to complementary levels while the \wordline is asserted to propagate these voltages to the internal nodes \stateq and \stateqb.
These nodes are connected to the drain and gate of the load \pfefets M2 and M4.
However, to ensure non-volatile storage, the write voltage is increased to \qty{4}{\volt} and applied with a longer pulse duration to reliably program the ferroelectric layer.
The programming schemes illustrated in \Cref{fig:fefet_write_schemes}, show that gate-drain and gate-source driven schemes produce asymmetric polarization strengths.
This asymmetry is quantified in \Cref{fig:fefet_write_iv_comp}, where the achievable memory window using these two schemes is smaller compared to conventional schemes.
With this scheme, the \pfefet cannot be fully polarized to the \hvt state.
However, this is not problematic because during nominal operation, the gate-source potential that the transistor in \hvt and \lvt will see are \qty{0}{\volt} and \qty{-1}{\volt} respectively, assuming maximum \qty{1}{\volt} power supply.
The transistor in the \hvt state is completely off at this potential while the transistor in \lvt is strongly conducting.

\begin{figure}
    \pgfmathsetmacro{\FigWidth}{0.625\columnwidth}
    \centering
    \footnotesize{}
    \begin{subfigure}[t]{0.5\columnwidth}
        \centering
        \includegraphics{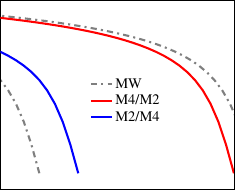}
        \caption{Pre-write state}
        \label{fig:idvg_write_a}
    \end{subfigure}%
    \begin{subfigure}[t]{0.5\columnwidth}
        \centering
        \includegraphics{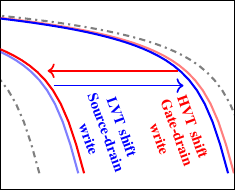}
        \caption{Opposite state write}
        \label{fig:idvg_write_b}
    \end{subfigure}
    
    \par\bigskip
    
    \begin{subfigure}[t]{0.5\columnwidth}
            \centering
            \includegraphics{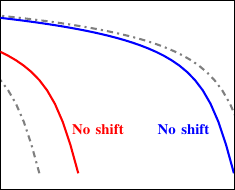}
            \caption{Same state write}
            \label{fig:idvg_write_c}
    \end{subfigure}%
    \begin{subfigure}[t]{0.5\columnwidth}
            \centering
            \includegraphics{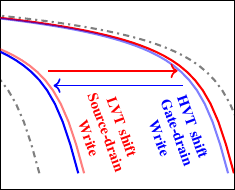}
            \caption{Opposite state write}
            \label{fig:idvg_write_d}
    \end{subfigure}%
    \caption{Write scheme in the context of \idvg shifts for the \pfefet transistors starting in (a) the pre-write state.
    (b) When writing the opposite state to what is stored inside the \sram, the polarization of the two transistors changes places.
    This is visible in terms of shifted \vth and \idvg curves.
    The semi-transparent red and blue \idvg curves in figures (a) and (c) show the previous polarization state before the \pfefet load transistors M2 and M4.
    Because of gate-drain writing schemes, the transistor does not get fully polarized to the \hvt state thus not reaching the maximal memory window.
    This memory window is enough to trigger the positive feedback of cross coupled inverters.
    When writing the same state, no change occurs in the polarization.}
    \label{fig:idvg_write}
    \vspace{-0.2cm}
\end{figure}

During this phase, the power supply is kept at \qty{4}{\volt}.
This prevents unintended conduction during the drain-write programming.
The load \pfefet with \qty{0}{\volt} at its gate is conductive; if there is a potential difference between its source and drain, then current can flow easily from the supply to the bitline, effectively creating a short.
Increasing the power supply to \qty{4}{\volt} avoids this issue.
This bias condition corresponds to the source and drain programming scheme in \Cref{fig:fefet_source_drain_write}, where the ferroelectric layer polarizes strongly to the \lvt state.

\begin{figure*}[t]
    \vspace{-0.3cm}
    \centering

    \begin{subfigure}[c]{\columnwidth}
        \centering
        \includegraphics{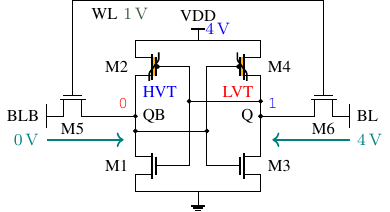}
        \subcaption{Write}
        \label{fig:nvsram_write}
    \end{subfigure}%
    \begin{subfigure}[c]{\columnwidth}
        \centering
        \includegraphics{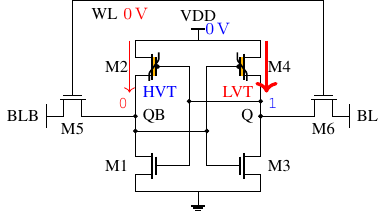}
        \subcaption{Power Off}
        \label{fig:nvsram_power_off}
    \end{subfigure}
    \vspace{0.3cm}

    \begin{subfigure}[c]{\columnwidth}
        \centering
        \includegraphics{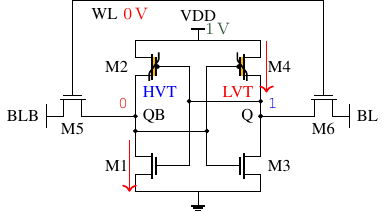}
        \subcaption{Restore}
        \label{fig:nvsram_restore}
    \end{subfigure}%
    \begin{subfigure}[c]{\columnwidth}
        \centering
        \includegraphics{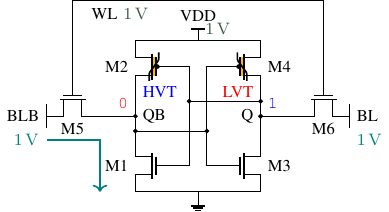}
        \subcaption{Read}
        \label{fig:nvsram_read}
    \end{subfigure}
    \caption{Proposed \nvsram voltage bias conditions and corresponding operations.
    (a) During write, high-voltage programming polarizes the load \pfefets into complementary \hvt/\lvt states.
    (b) Polarization of \pfefets is preserved during power-off and internal storage nodes \stateq and \stateqb collapse without data loss.
    (c) Upon power ramp-up, the current imbalance between the polarized load devices biases the cross-coupled inverters and re-establishes the potential on the storage nodes.
    (d) Conventional 6T SRAM read operations where the precharged bitlines are discharged via the pull-down NFET transistors.}
    \label{fig:nvsram_circuit_operation}
\end{figure*}

If the incoming data matches the existing state, no polarization switching occurs (\Cref{fig:idvg_write}).
If the data is different, the \pfefet experiencing $\vg=\qty{4}{\volt}$, $\vs=\qty{4}{\volt}$ and $\vd=\qty{0}{\volt}$, switches from the \lvt state to an \hvt state.
Simultaneously, the other \pfefet shifts towards the \lvt state due to the potential difference across its gate, source and drain; $\vg=\qty{0}{\volt}$, $\vs=\qty{4}{\volt}$, and $\vd=\qty{4}{\volt}$.
\Cref{fig:idvg_write} visualizes the state-dependent \vth evolution of M2 and M4 during write operation. When writing the opposite state, the polarization of the two load \pfefets swaps, producing complementary \vth shifts.
The voltage on different nodes of the \nvsram cell for the case of logic \logicone write are shown in \Cref{fig:nvsram_write}.
After the polarization pulse is applied to \bitline and \blb, the \wordline is de-asserted.
The voltages at nodes \stateq and \stateqb return to their nominal values of \qty{1}{\volt}.
Note that during this hold state, the voltage values on \stateq and \stateqb, despite being smaller than the voltages required for polarization, reinforce the polarization of the \fefet transistors.
This guarantees stable data and prevents random bit flips.

\subsection{Power-Off and Restore}
The non-volatility of the proposed \nvsram, due to the polarized load \pfefets, allows for safe power-down of the entire \sram array without data loss.
This can be easily performed in a one-step approach where the power supply \vdd is pulled to ground. \Cref{fig:nvsram_power_off} shows the potential levels on the relevant nodes when \logicone is stored. The internal node holding \qty{1}{\volt} is pulled close to ground by the \lvt transistor during this stage. The polarization of the \pfefets is not disturbed during power off.
 
During the restore or power-on operation, the power supply is ramped up to the operation voltage of \qty{1}{\volt}.
The ramp-up of \vdd cannot be arbitrarily fast.
The restore process relies on the current imbalance between the \hvt and \lvt polarized \pfefets, which must develop before the cross-coupled inverter pair enters strong positive feedback.
As shown in \Cref{fig:fefet_write_iv_comp}, a current imbalance exists across the gate-source voltage range from \qty{-1}{\volt} to \qty{0}{\volt} since the maximum supply voltage \vdd is \qty{1}{\volt}.  However, the amplitude of this current differential depends on the voltage potential on the storage nodes \stateq and \stateqb before the restore. Due to process variation and noise from the power-up ramp, the value of the current imbalance might be very small, which might force the circuit to latch into an unstable state and in the worst case scenario a short circuit state. The soft-start via the slow power-up help prevent this problem by limiting noise and allowing the positive feedback to place the storage nodes to the expected values incrementally.

In addition to process variation, residual charge can accumulate on nodes \stateq and \stateqb due to the surrounding fields by interconnects in the proximity of the cell, potentially placing them outside the expected \qty{-1}{\volt} to \qty{0}{\volt} range. This and process variation might lead to a sign change in the differential current between the two load \pfefets.
This may lead to incorrect data latching during restore.
In this case, the restore sequence can be extended to a two-phase approach where first the \bitline and \blb are pulled to \qty{0}{\volt} with the \wordline asserted.
This ensures that both load \pfefets see the same potential on their gate, drain and source during power-up.
The transistor polarized to the \lvt state is able to drive more current in the next phase of \vdd ramp-up to which its drain is connected to.
Because one \pfefet conducts more current, one of the nodes \stateq and \stateqb is pulled faster towards the supply voltage.
At a voltage level determined by the polarization state, the positive feedback cross-coupled inverter takes over. As shown in \Cref{fig:nvsram_restore}, for the case when \logicone is stored, due to the positive feedback M1 pulls down the node \stateqb to ground while the load \pfefet M4 pushes charges to the node \stateq.
The \nvsram cell then latches to the correct data state.

\begin{figure*}[t]
    \pgfmathsetmacro{\plotWidth}{4.3cm}
    \centering

    \begin{subfigure}[c]{0.3\textwidth}
        \centering
        \includegraphics{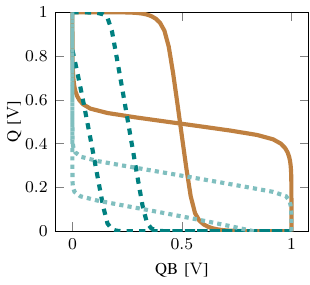}
        \subcaption{\acrshort{hsnm}}
        \label{fig:HSNM}
    \end{subfigure}%
    \begin{subfigure}[c]{0.3\textwidth}
        \centering
        \includegraphics{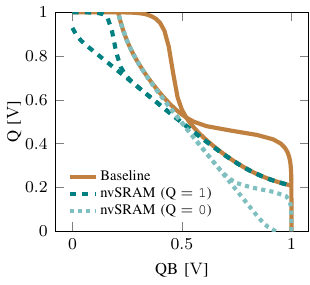}
        \subcaption{\acrshort{rsnm}}
        \label{fig:RSNM}
    \end{subfigure}%
    \begin{subfigure}[c]{0.3\textwidth}
        \centering
        \includegraphics{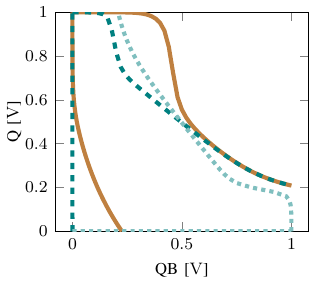}
        \subcaption{\acrshort{wsnm}}
        \label{fig:WSNM}
    \end{subfigure}
    \caption{Comparison of \acrshortpl{snm} for 6T \sram and proposed \nvsram.
    Here, it should be noted that for our proposed \nvsram, the \acrshort{hsnm} and \acrshort{rsnm} values look smaller than in the 6T \sram.
    However, since the cell is monostable with only one visible lobe, the smaller \acrshort{hsnm} and \acrshort{rsnm} will not cause any problems. WSNM is shown only for comparison, as the writting for the proposed design is done at elevated \bitline/\blb and \vdd voltages.}
    \label{fig:SNM}
    \vspace{-0.2cm}
\end{figure*}

\subsection{Read Operation}
The read operation of the proposed design is identical to that of the baseline 6T \sram.
Both bitlines are precharged and \wordline is asserted.
The stored data causes a differential discharge of \bitline and \blb, which is detected by a sense amplifier.
\Cref{fig:nvsram_read} shows the biasing of the \nvsram cell during a logic \logicone read operation.
Read operations are performed at $\vdd = \qty{1}{\volt}$.
Since the read current path flows through the pull-down network, the load \pfefets do not participate in the discharge path. This preserves conventional read performance.

\subsection{Hold Operation}
The hold operation functions identically to the baseline 6T \sram with the data actively retained through the cross-coupled inverters.
However, the primary advantage of the proposed \nvsram is the ability to power down the array during idle periods, reducing standby energy consumption to effectively zero. 

\section[Simulation Results of Our Proposed nvSRAM Design]{Simulation Results of\\Our Proposed nvSRAM Design}
\label{sec:setup}

The proposed \nvsram design is evaluated through \spice simulations in Cadence Virtuoso using a commercial \qty{28}{nm} technology.
A bitline capacitance of \qty{17}{fF} is assumed, corresponding to a column of 64 cells \cite{UCB}.

\subsection{Static Noise Margin (SNM) and Monostability}
\snm is used to evaluate the resilience of the \sram against voltage fluctuations on the internal storage nodes under hold, read and write conditions. \Gls{hsnm} reflects the inverter strength, \rsnm depends on the pull-down to access transistor ratio, and \wsnm is governed by the access to pull-down strength ratio \cite{SNM_ratio}.
In the baseline 6T \sram, the cross-coupled inverters are identical, resulting in a symmetric butterfly curve with two stable lobes for \hsnm and \rsnm. 

\begin{figure}
    \centering
    \begin{subfigure}[t]{\columnwidth/2}
        \centering
        \includegraphics{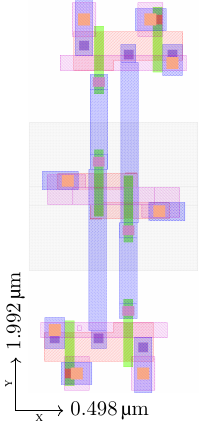}
        \subcaption{\nvsram}
        \label{fig:nvsram_layout}
    \end{subfigure}%
    \begin{subfigure}[t]{\columnwidth/2}
        \centering
        \includegraphics{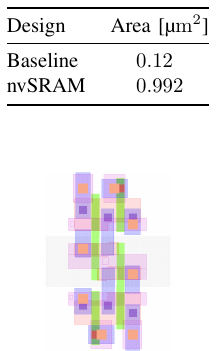}
        \subcaption{Baseline \sram layout}
        \label{fig:baseline_sram_layout}
    \end{subfigure}
    \caption{\sram layout comparison.
    The area increase for the minimum-size, DRC-compliant \nvsram is \qty{8.3}{\times} that of the baseline \sram.
    This is mainly due to the spacing between the normal \nmos gate and the \pfefet gate.
    The gate pitch between two gates of the same type is also increased to accommodate DRC rules.}
    \label{fig:layout_comparison}
    \vspace{-0.4cm}
\end{figure}

In contrast, our proposed \nvsram exhibits monostable behaviour.
Because the \pfefets are programmed into complementary \vth states (one \hvt and one \lvt), the inverters are intentionally non-identical.
Although this departs from conventional \sram design practice, it is essential for enabling non-volatility.
\Cref{fig:HSNM} and \Cref{fig:RSNM} show the \hsnm and \rsnm for the proposed \nvsram design.
The butterfly curves in these figures show only a single lobe per stored state.
This asymmetric \snm characteristic directly reflects the intentional strength imbalance introduced by the polarized \pfefet load transistors.
While write operations occur at \qty{4}{\volt}, where noise is negligible, the \qty{1}{\volt} \wsnm is shown in \Cref{fig:WSNM}.
\wsnm for the \nvsram remains similar to the baseline 6T \sram.
Overall, \snm evaluation for \nvsram requires separate \hsnm and \rsnm curves for each polarization state to fully characterize the monostable behaviour.

\begin{table}[t]
    \centering
    \caption{Comparison of latency for baseline and proposed \nvsram}
    \renewcommand{\arraystretch}{1.2}
    \begin{tabular}{@{}l
    S[table-format=2.0, table-number-alignment=center]@{}}
    \toprule
    Metric            & {Read Latency [\unit{ps}]}   \\
    \midrule
    Baseline & 70 \\ 
    \nvsram & 72 \\ 
    \bottomrule
    \end{tabular}
    \label{tab:nvsram_to_baseline_comp}
    \vspace{-0.2cm}
\end{table}

\subsection{Latency Analysis}
Read latency is evaluated and compared with the baseline 6T \sram.
The results are displayed in \cref{tab:nvsram_to_baseline_comp}.
We achieve a \qty{72}{ps} read latency compared to the \qty{70}{ps} latency of the baseline \sram cell.
The read latency is defined as the time required to develop a \qty{100}{\milli\volt} potential difference between \bitline and \blb after the \wordline is asserted.
Because the read path is dominated by the access and pull-down transistors, which remain unchanged in our \nvsram, the read latency remains comparable to that of the baseline 6T \sram.
This confirms that replacing the \pmos load devices with \pfefets does not degrade the critical read path of the memory cell.

In contrast, the write operation of the proposed \nvsram design rely on \fefet programming using \qty{4}{\volt} pulses with a \qty{10}{ns} duration \cite{dunkel2017fefet_22nm, mulasmanovic2019gf_fefet_endurance}.
Consequently, the \nvsram write latency is significantly higher than the baseline design in the similar node.
However, in many memory-centric applications, the number of reads exceeds that of writes \cite{read_write_ratio}.

\subsection{Area Overhead}
The proposed \nvsram achieves non-volatility without additional transistors or capacitors, resulting in improved architectural area efficiency compared to state-of-the-art \fefet-based designs.
The layout of the baseline 6T \sram and our proposed \nvsram design is shown in \cref{fig:layout_comparison}.
Although the transistor count is identical, the physical layout of the \nvsram is \qty{8.3}{\times} larger than the baseline design.
This overhead is attributed to the design-rule constraints associated with the \fefet layer.
Since the pull-down \nmos transistors in the \nvsram design are conventional \mosfet transistors, gate sharing is not possible.
Nevertheless, the area of our design remains smaller than existing \nvsrams that require additional \fefet transistors or capacitors subject to the same gate-to-gate spacing.
At the array level, our design does not require additional control peripherals, which are commonly needed in state-of-the-art \nvsram architectures.

\section[Measurements of Our Fabricated nvSRAM Design]{Measurements of Our\\Fabricated nvSRAM Design}
\label{sec:measurements}

\subsection{Ferroelectric FET Device Measurements}
\label{sec:device_measurements}

\begin{figure}[t]
  \centering
  \includegraphics[width=\columnwidth]{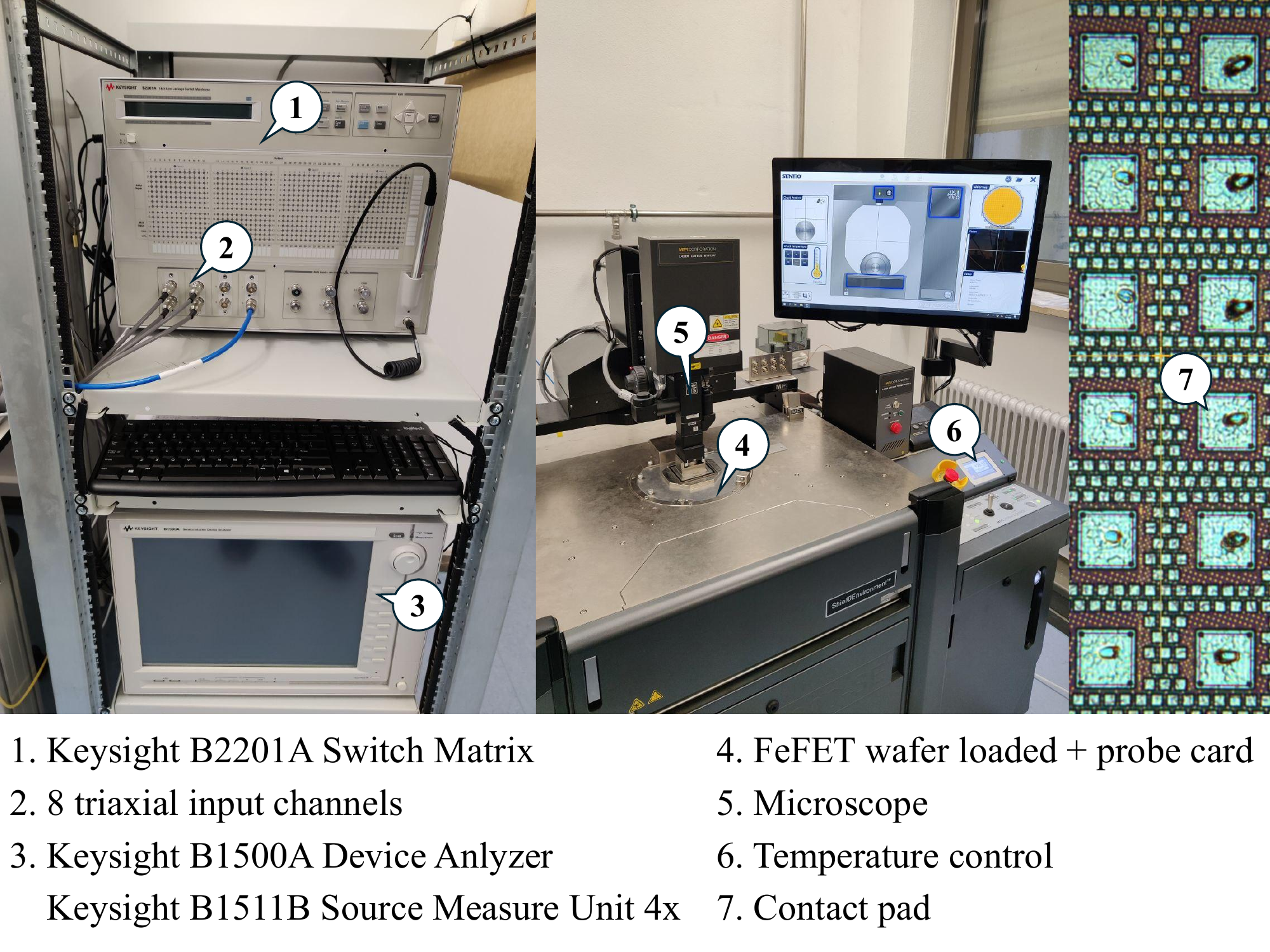}
  \caption{Setup for measuring our fabricated \fefet devices and die chips using the MPI TS3000-SE \qty{300}{mm} wafer prober.}
  \label{fig:measurement_setup}
\end{figure}

\begin{figure}[t]
    \centering
    \begin{subfigure}{\columnwidth}
        \phantomsubcaption
        \label{fig:nmos}
        \phantomsubcaption
        \label{fig:pmos}
    \end{subfigure}
    \input{paper-gnuplottex-fig1.tex}
    \caption{Transfer characteristic of \fefet devices fabricated in bulk \qty{28}{nm} with cycle variation.
    (a) \idvg loop of \nfefet with a channel dimension of \chdimsqr{\qty{500}{\nano\meter}}.
    (b) \idvg loop of \pfefet with a channel dimension of \chdimsqr{\qty{500}{\nano\meter}}.
    All \idvg characteristic extractions have been performed at a \vds of \qty{50}{mV} and polarized with \qty{-4}{V} for \hvt and \qty[retain-explicit-plus]{+4.5}{V} for \lvt (opposite signs for \pfefet).}
    \label{fig:iv_measurements}
    \vspace{-0.2cm}
\end{figure}

We used the MPI TS3000-SE wafer probe station for the device characterization and die measurements (\cref{fig:measurement_setup}).
Loaded on the thermal-controlled chuck maintained at \qty{300}{K} is the \qty{300}{mm} wafer with the devices and circuits.
For electrical connectivity, a probe card is designed with \num{12} needles contacting \num{12} pads in parallel.
The triaxial cable harness running from the probe card holder is connected to the \num{12} output channels of the Keysight B2201A switch matrix.
With the switch matrix, the \num{8} input channels can be connected to the output channels in any configuration.
This makes it easy to change devices in the pad column currently touched by the needles.
At the end of the signal path is the Keysight B1500A semiconductor device analyzer mainframe.
We use the \num{4} installed Keysight B1511B \smus connected to the \num{8} input channels of the switch matrix via triaxial cables to drive the biases and measure currents.
With the shipped Keysight Expert Group Plus software, we create the application tests that control measurement procedures on the B1500.
\cref{fig:measurement_setup} illustrates a detailed overview of our measurement setup.

To characterize the device \idvg transfer characteristics, we use the switch matrix to connect the \smus to the \dut terminals.
In case of \pfefet, the \dut is first reset to the \hvt state by using a \qty[retain-explicit-plus]{+4}{V} gate pulse.
To polarize to the \lvt state, a sufficiently long \qty{-4.5}{V} pulse is used.
After the polarization, the gate bias is swept from \qtyrange{1}{-1.5}{V} in \qty{50}{mV} steps while measuring the \ids current.
We use a reverse sweep (i.e., going from positive to negative) for the \pfefet to mitigate polarization towards the \lvt state during the \idvg extraction sweep.
To ensure sufficient de-trapping, \qty{2}{s} wait time is inserted before the gate sweep.
The source and body terminal are biased with \qty{0.8}{V}, and to establish a \qty{50}{mV} \vds difference, the drain terminal is biased with \qty{0.75}{V}.
To collect cycle-to-cycle variation data, the reset-and-polarization scheme is repeated \num{10} times for both the \lvt and \hvt states.
To measure the \nfefet, the \dut is reset to the \hvt state with a \qty{-4}{V} gate pulse and polarized to the \lvt state with a \qty[retain-explicit-plus]{+4.5}{V} gate pulse.
The source and body terminal are grounded while the drain terminal is biased with \qty{50}{mV}.
For the \idvg sweep we go from \qtyrange{-1}{1.5}{V} in \qty{50}{mV} steps.

We have designed and placed a small array with different channel dimensions for both \nfefet and \pfefet on the die.
To measure it, we use the switch matrix in combination with the probe card to connect to the targeted \dut's corresponding terminals without recabling.
The measurement results for both \nfefet and \pfefet with a channel dimension of \chdimsqr{\qty{500}{\nano\meter}} are shown in \cref{fig:iv_measurements}.

\subsection{Silicon Validation of Our nvSRAM Design}
\label{sec:nvsram_measurements}

\begin{figure}[t]
    \centering
    \includegraphics{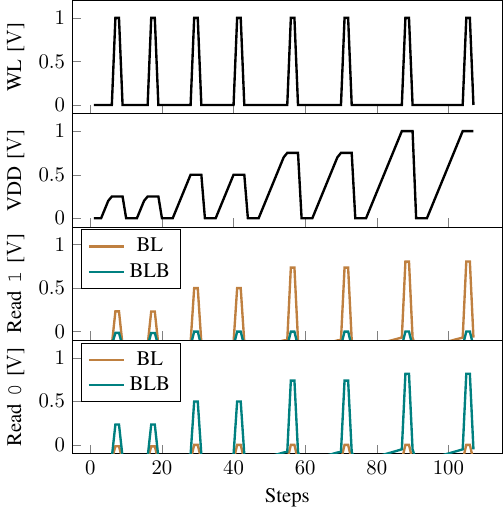}
    \caption{Measurment results of our proposed \nvsram design fabricated in \qty{28}{nm} technology.
    The plot shows the biases applied to \wordline and \vdd and the measured voltages of \bitline and \blb for the read \code{1} and \code{0} case.
    As \smus are employed to operate and measure the circuit, the x-axis is \emph{not} transient, but shows the steps in the order we step through them.
    To restore, we ramp \vdd to the target voltage in multiple steps to allow the cell to latch into the correct state.
    Following the restore and before enabling \wordline, we hold \vdd at the target voltage for one step.
    To test the reliability of the restore procedure, we use several target \vdd{}s (\qtylist{0.25;0.5;0.75;1.0}{V}) with two tests per target \vdd.
    The results for the read \code{1} and \code{0} cases show that the \nvsram cell latches to the correct states at each restore.}
    \label{fig:measurement_nvsram}
    \vspace{-0.2cm}
\end{figure}

For silicon verification, we have taped-out our proposed \nvsram design in \qty{28}{nm} technology.
Our \pfefet{}s dimensions are \qty{8}{um} for the width and \qty{30}{nm} for length.
To perform the measurements, we use the same measurement setup as discussed in \cref{sec:device_measurements} and shown in \cref{fig:measurement_setup}.
Using the B1500A, we drive the voltage biases to operate our \nvsram.
\cref{fig:measurement_nvsram} shows the measured results for the two cases of read \code{1} and \code{0}.
In the read operation, the \nvsram cell is driven by applying voltage biases to \wordline and \vdd.
Using the \smus of the B1500A, the voltages of \bitline and \blb are measured.
As we use \smus to perform the measurements, there is no strict notion of time (e.g., like a waveform recorded with an oscilloscope), but rather measurement points that are stepped through in consecutive sequence.
Hence, \cref{fig:measurement_nvsram} is \emph{not} a transient plot.
It can be understood as a plot of consecutive measurements of the stable operating point at each step, with transitions between steps.

\begin{figure}[t]
    \centering
    \input{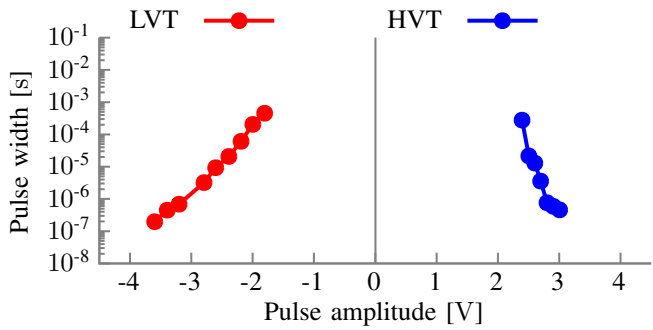}
    \caption{Halid plot of \qty{28}{nm} \pfefet linking the switching dynamics of amplitude and pulse width.
    The measurements demonstrate that with a voltage applied to the \fefet gate of up to \qty{1}{V}, no switching in the ferroelectric polarization can happen (i.e., no disturbance).
    Data from \cite{kleimaier2021p_fefet}.}
    \label{fig:halid_plot}
    \vspace{-0.375cm}
\end{figure}

To restore the \sram state, we ramp \vdd to the target voltage in \qty{100}{mV} steps.
The gradual \vdd ramp increases the probability that the \nvsram latches to the correct state during restore.
After the ramp-up and before we enable the \wordline, we add one step to hold \vdd at the target voltage.
This is to let the circuit settle in the state and ensure all signals are stable.
Following that, we enable the \wordline and the internal values of the \nvsram cell are forwarded to the \bitline and \blb.
We power off the \nvsram cell in between each read to prepare it for the next restore cycle.
To test the reliability of the restore process, we repeat it two times per target \vdd, and test several target \vdd{}s (\qtylist{0.25;0.5;0.75;1.0}{V}).
When connecting the measurement equipment, significant capacitance and noise are added.
To counteract that, we avoid unwanted accumulation of charges on \bitline and \blb by continuously drawing a current of \qty{1}{\micro\ampere}.
Additionally, we draw current to establish a low voltage level, allowing us to measure the strength of the \pfefet, the key transistor in our design.
Due to the constant current draw on \bitline and \blb, the voltage is negative in between the cell readouts.
Still, the \nvsram cell is strong enough to establish the \qty{\approx0}{V} and targeted \vdd voltage level on \bitline and \blb, depending on the stored state.
As shown in \cref{fig:measurement_nvsram}, the \nvsram cell latches to the correct state every time, validating our proposed \nvsram cell design.

\subsection{FeFET Reliability Discussion}
\label{sec:reliability_discussion}

\textbf{Endurance:}
The endurance of \fefets is a concern in this use case as \sram itself offers a high endurance of \num{>e15}.
Endurance measurements of \qty{28}{nm} \fefet devices shows an endurance of \num{e5} for \nfefet and \num{e4} for \pfefet \cite{kleimaier2024p_n_type_fefet_endurance}.
This means that the non-volatile storage of data in the \pfefet{}s can not be used as frequently as changing the state of the underlying \sram cell.
If needed, the endurance can be increased by utilizing lower programming voltages (e.g., \qty{3.3}{V} instead of \qty{4.5}{V}) \cite{sharma2020intel_fefet_endurance}.

\textbf{Disturbance:} During the hold of the \nvsram cell, one of the \pfefet transistors has a constant gate bias applied.
The polarization process also depends on the pulse width, i.e., the duration of the applied pulse.
Hence, a weaker electric field applied for a long duration may also switch the domain's polarization, raising concerns about potential disturbance.
However, the halid plot based on \pfefet \qty{28}{nm} measurements shown in \cref{fig:halid_plot} links the switching dynamics of pulse amplitude and width.
Stronger amplitudes need shorter pulses to shift the polarization from, e.g., \hvt to \qty{50}{\percent} \lvt.
The data shows, for voltages below \qty{2}{V}, long pulses (\qty{>100}{s}) are needed to change the polarization, if it is changing at all.
Therefore, with a \vdd of \qty{1}{V}, the electric field from the gate and drain node is small enough not to cause disturbance in the ferroelectric state.
Additionally, as stated earlier, the bias of \stateq and \stateqb during the hold operation positively reinforces the polarization.

\textbf{Retention:} To serve as a \nvm, the retention is an important metric.
It estimates the time over which the polarization state is retained.
In our use case, this means for how long the \nvsram cell can remain powered off, hold the state in the \pfefets, and latch to the correct state when restored. 
For the used \qty{28}{nm} process, the retention is extrapolated to be \num{10} years at a stress temperature of \qty{105}{\celsius} \cite{trentzsch2016hkmg28nm_fefet}.
With it, our proposed \nvsram cell design can retain the state over its entire estimated lifetime. 

\section{Conclusion}
Utilizing the recently demonstrated \pfefet we have designed a novel \nvsram cell without the need for additional transistors.
By replacing the pull-up \pmos transistors with \pfefets, we achieve a non-volatile cell that offers a reliable power-down and restore.
Using simulation, we tested the proposed concept and analyzed its reliability using \snm.
For silicon verification, the design was taped-out using a \qty{28}{nm} process and measured.
The measurements confirm the proof of concept and successful power-up using different supply voltages.
From the layout of our proposed \nvsram cell, the area footprint consumed is \qty{0.99}{\micro\meter\squared}.

\section*{Acknowledgment}
We thank Halid Mulaosmanovic, Stefan Dünkel, and Sven Beyer from GlobalFoundries for fabricating our designs, and Anirban Kar for his insights and discussions that contributed to the development of the idea. 

\bibliographystyle{IEEEtran}
\bibliography{refs}

\begin{IEEEbiography}[{\includegraphics[width=1in,height=1.25in,clip,keepaspectratio]{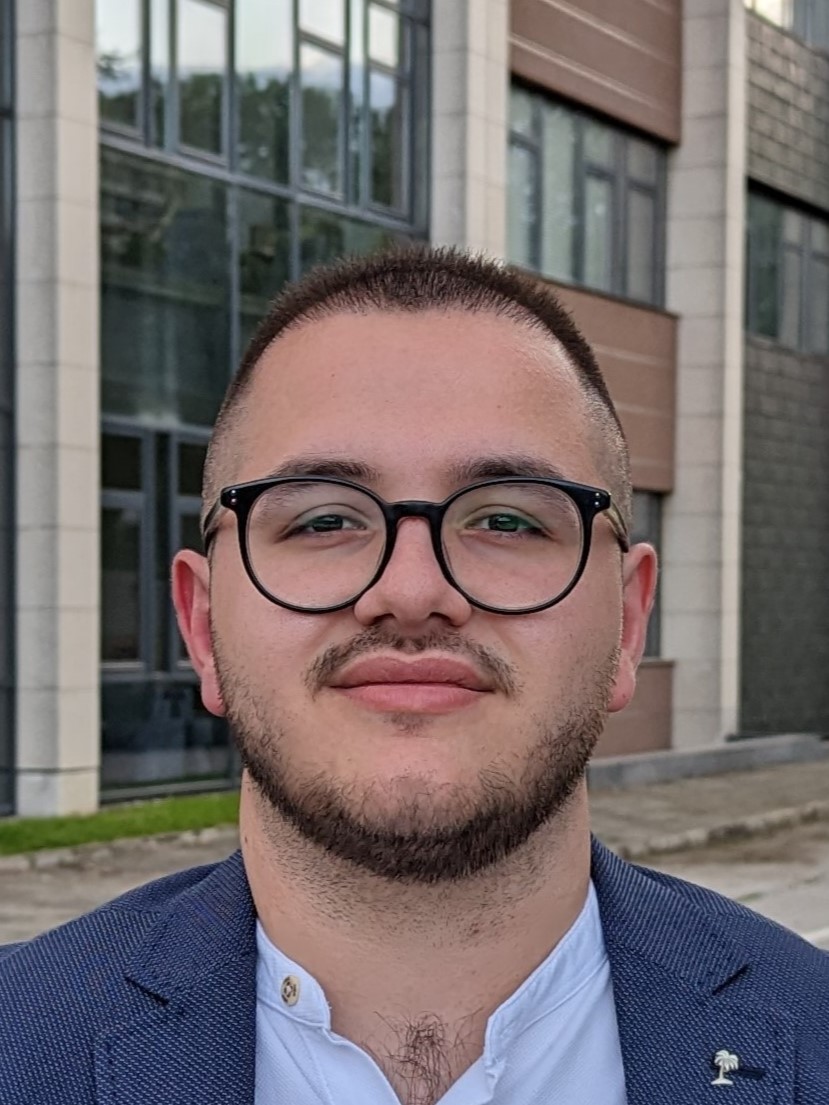}}]%
{Albi Mema} (M'24) is currently a Ph.D. candidate at the AI Processor Design (AI-PRO) chair at the Technical University of Munich. He earned his Bachelor degree in Electrical and Computer Engineering Major and Intelligent Mobile Systems (’19) at Constructor University (formerly named Jacobs University Bremen). He earned his Masters degree in INFOTECH at the University of Stuttgart ('22). His technical expertise lies in the area of analog and digital circuit and chip design. His main research interests are in in-memory, analog, and neuromorphic computing, ferroelectic devices, and advanced packaging.
\end{IEEEbiography}

\begin{IEEEbiography}[{\includegraphics[width=1in,height=1.25in,clip,keepaspectratio]{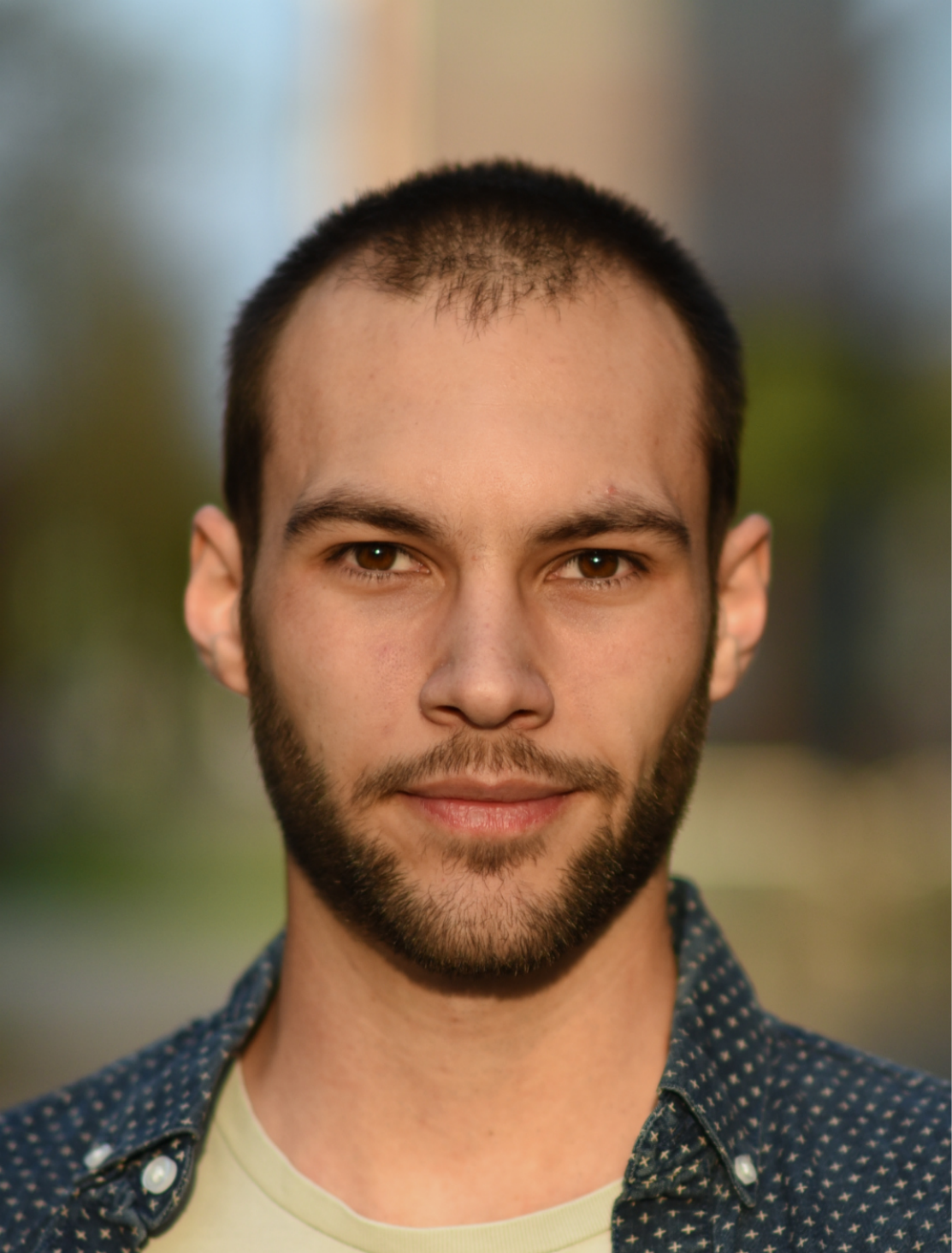}}]%
{Simon Thomann} (M'22) earned his degrees in Computer Science, Master in 2022 as well as Bachelor in 2019, at Karlsruhe Institute of Technology (KIT), Germany.
He started his Ph.D. at the University of Stuttgart in 2022 and has been continuing it since 2023 at the Chair of AI Processor Design within the Technical University of Munich (TUM).
His research interests range from device physics to the system level.
His special interest lies in circuit design, emerging technologies, and cross-layer reliability modeling from device to circuit level.
\end{IEEEbiography}

\begin{IEEEbiography}[{\includegraphics[width=1in,height=1.25in,clip,keepaspectratio]{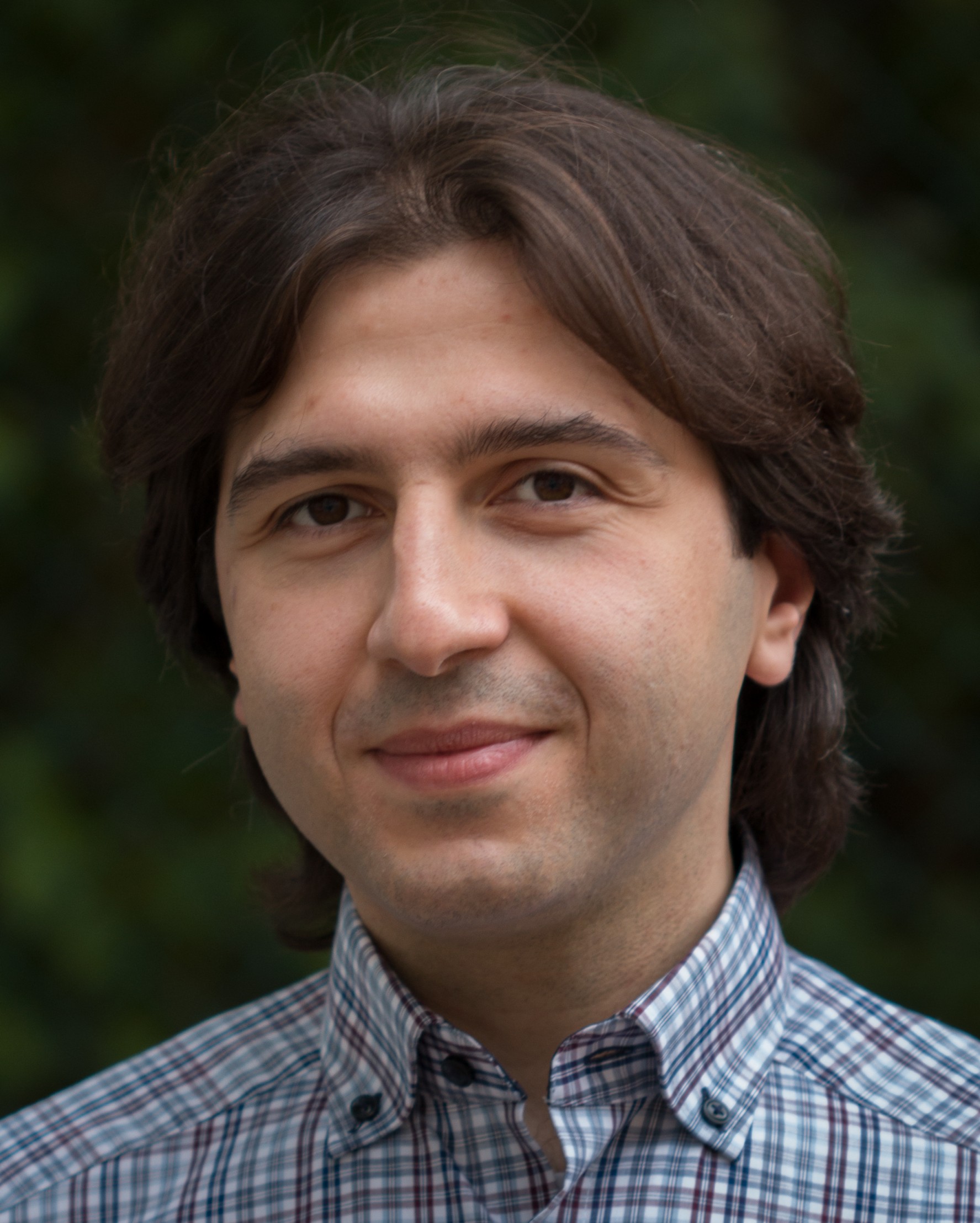}}] {Hussam Amrouch}(S'11-M'15) is Professor heading the Chair of AI Processor Design within the Technical University of Munich (TUM). He is the Founding Director of the Munich Advanced-Technology Center for High-Tech AI Chips (MACHT-AI). He is the head of the Brain-inspired Computing within the Munich Institute of Robotics and Machine Intelligence in Germany and is also the head of the Semiconductor Test and Reliability  within the University of Stuttgart, Germany. Prior to that, he was a Research Group Leader at the Karlsruhe Institute of Technology (KIT) where he was leading the research efforts in building dependable embedded systems. He currently serves as Editor at the Nature Scientific Reports Journal. He received his Ph.D. degree with the highest distinction (Summa cum laude) from KIT in 2015. His main research interests are design for reliability and testing from device physics to systems, machine learning for CAD, HW security, approximate computing, and emerging technologies with a special focus on ferroelectric devices. He holds 10x HiPEAC Paper Awards and three best paper nominations at top EDA conferences: DAC'16, DAC'17 and DATE'17 for his work on reliability. He has served in the technical program committees of many major EDA conferences such as DAC, ASP-DAC, ICCAD, etc., and as a reviewer in many top journals like Nature Electronics, T-ED, TCAS-I, TVLSI, TCAD, TC, etc. He has more than $310$ publications in multidisciplinary research areas (including over $130$ journals) across the entire computing stack, starting from semiconductor physics to circuit design all the way up to computer-aided design and computer architecture. His research in HW security and reliability have been funded by the German Research Foundation (DFG), Advantest Corporation, and the U.S. Office of Naval Research (ONR). 
\end{IEEEbiography}

\end{document}